\documentclass[12pt,twoside]{article}

\usepackage{fleqn,espcrc1}
\usepackage{graphicx}

\title{\center{Information on antiprotonic atoms and the nuclear periphery from
the PS209 experiment}}
\author{A.~Trzci\'nska\address[SLCJ]{Heavy Ion Laboratory, Warsaw 
University, PL-02-093 Warsaw, Poland}, 
J.~Jastrz\c{e}bski\addressmark[SLCJ], 
T.~Czosnyka\addressmark[SLCJ], 
T.~von~Egidy\address[TUM]{Physik-Department, 
Technische Universit\"at M\"unchen, D-85747 Garching, Germany},
K.~Gulda\address[IFD]{Institute of Experimental Physics,
Warsaw University, PL-00-681, Warsaw, Poland},
F.~J.~Hartmann\addressmark[TUM], 
J.~Iwanicki\addressmark[SLCJ], 
B.~Ketzer\addressmark[TUM], 
M.~Kisieli\'nski\addressmark[SLCJ], 
B.~K{\l}os\address{Institute of Physics, University of Silesia, 
PL-40-007 Katowice, Poland}, 
W.~Kurcewicz\addressmark[IFD]
P.~Lubi\'nski\addressmark[SLCJ], 
P.~J.~Napiorkowski\addressmark[SLCJ],
L.~Pie\'nkowski\addressmark[SLCJ], 
R.~Schmidt\addressmark[TUM],
E.~Widmann\address{CERN, CH-1211 Geneva 23, Switzerland} }

\begin{document}
\maketitle

The PS209 experiment was run during two three-week periods in 1995
and 1996. The antiproton beam momenta were 412~MeV/c and 310~MeV/c
in the first period and 106~MeV/c in the second one. A fraction of
the beam time was used for the continuation of the radiochemical
experiments aiming at the determination of the peripheral neutron
to proton density ratio for 19 medium mass and heavy nuclei. The
method, proposed by our collaboration some years ago~\cite{jast93},
consists in the study of the
annihilation residues with the mass number one unit smaller than the
target mass $A_t$. When  both $A_t -1$ products (i.e. those with
proton number $Z_t -1$ and those with neutron number $N_t -1$) 
are radioactive, their relative yields
after antiproton annihilation are easily determined by 
standard nuclear spectroscopy methods. They are directly related
to the proton and neutron densities at the annihilation site.
The radial distance of the most probable value of the 
annihilation site for events leading to $A_t -1$ products is
obtained from calculations~\cite{wyc96} as $R_{1/2} + (2.5 \pm 0.5)$\,fm,
almost independent of the atomic number of the target. 
If one assumes that the peripheral proton
distributions of nuclei are now rather well determined using
electromagnetically interacting probes, our experiment may be
considered as a new way to evaluate the neutron
distributions.
In Ref.~\cite{lub94} the so-called halo factor $f_{halo}$ was introduced
as a measure for the neutron-over-proton density ratio.
This halo factor transforms
the measured yield of the $N_t -1$ to $Z_t -1$ nuclei into 
the corresponding normalised (by $Z/N$ factor)
density ratio of neutrons to protons in the target
nucleus at the radial distance of the annihilation site.
In Refs.~\cite{lub98} and~\cite{sch99} 
the halo factor was presented as a function of the target
neutron binding energy $B_n$.
A strong negative correlation was observed. For target nuclei with
$B_n < 9.5$~MeV clear evidence for a neutron rich periphery
was obtained.
\begin{figure}[h]
\begin{center}
\includegraphics[width=0.75\textwidth]{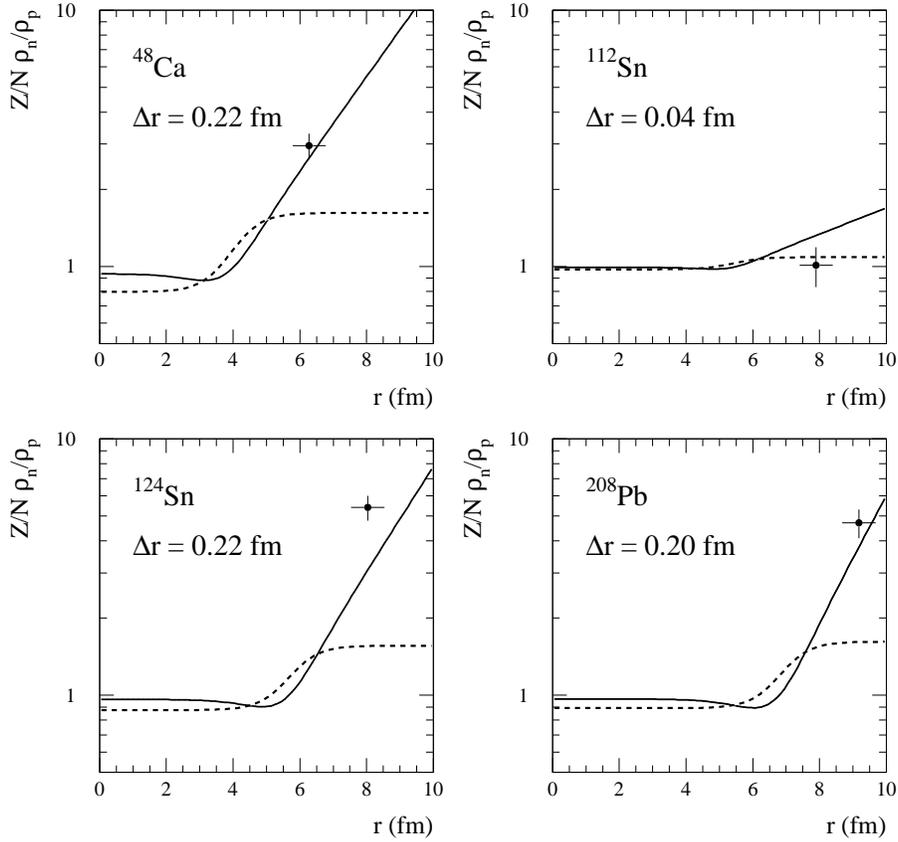}
\end{center}
\caption{The normalized neutron to proton density ratio deduced from the 
experimental $\Delta r_{np}$ for $^{48}\mbox{Ca}$~\cite{bat89}, 
 $^{112}\mbox{Sn}$ (extrapolated from Ref.~\cite{kra99}), 
$^{124}\mbox{Sn}$~\cite{kra94,kra99,bat89} 
and $^{208}\mbox{Pb}$~\cite{kra94,sta94}  nuclei. 
Crosses indicate the density ratio deduced from our radiochemical
experiments (interpolated value for $^{208}\mbox{Pb}$) presented at the
most probable annihilation site under the assumption of
$R = \frac{Im \, a(\overline{{\rm p}}{\rm n})}
{Im \, a(\overline{{\rm p}}{\rm p})}=0.63$. 
Solid line: $c_n=c_p$ (''neutron halo'' model), 
dashed line $a_n=a_p$ (''neutron skin'' model).}
\label{1}
\end{figure}

With the information on peripheral neutron densities from antiproton
annihilation, one is tempted to compare them with experiments
determining the difference $\Delta r_{np}$ between neutron and proton 
mean square radii.
Care, however, is necessary, as the peripheral densities are probed
by antiprotons at distances much larger than the root mean square 
radii. Keeping this limitation in mind, we nevertheless
compared the neutron over proton densities for a number of nuclei
in which recent~\cite{kra94,kra99,sta94} or older~\cite{bat89} 
$\Delta r_{np}$
data exist. Figure~\ref{1} gives the example for such a comparison.
In preparing this figure the experimentally determined $\Delta r_{np}$
values were first used to obtain the neutron $rms$ radius from the relation:
$ \langle r_n^2 \rangle ^{1/2} = \langle r_p^2 \rangle ^{1/2}
+ \Delta r_{np}$ where $\langle r_p^2 \rangle ^{1/2}$ was taken
from the recent tabulation~\cite{fri95} after correction for the proton
charge radius. 
The same tabulation gives the two-parameter Fermi (2pF)
charge distributions. These distributions
were converted to point proton distributions~\cite{ose90}.
The parameters for bare neutron distributions were obtained from
the relation $\langle r_n^2 \rangle  \approx \frac{3}{5} \, c_n^2 +
\frac{7}{5} \pi^2 a_n ^2$ assuming either  $c_n =c_p$ or $a_n = a_p$. 
These two cases are shown in Fig.~\ref{1}. As can be seen from this figure
our radiochemical data are clearly in favour of interpreting the
$\Delta r_{np}$ by the increase of the neutron surface 
diffuseness rather than the increase of the neutron half--density radius.

\begin{figure}[h]
\begin{center}
\includegraphics[width=0.55\textwidth]{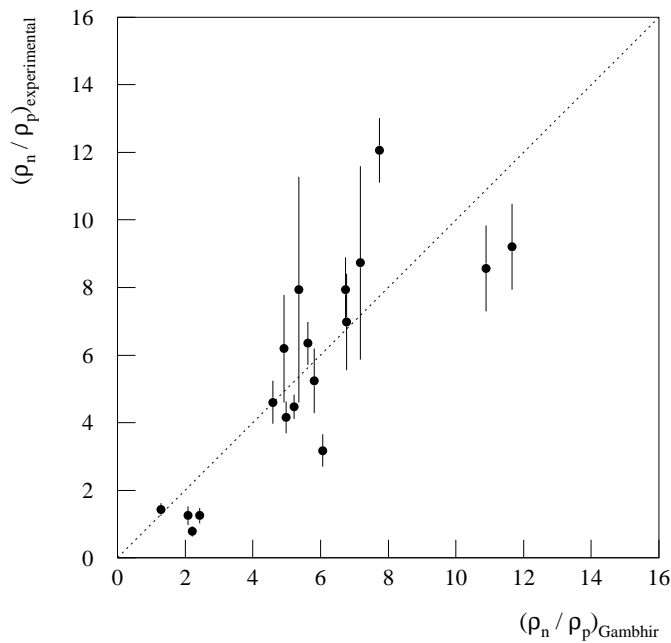}
\end{center}
\caption{Comparison of the experimental neutron to proton density
ratio at a radial distance of $R_{1/2} + 2.5$~fm with the result 
of the semiempirical Gambhir approach.}
\label{2}
\end{figure}

The ratios of neutron over proton density determined by the radiochemical
method were also compared with the semiphenomenological approach of
Gambhir and~Patil~\cite{gam86}.
Figure~\ref{2} compares the experimental density ratios with those
deduced from the Gambhir prescription. The good overall agreement,
together with the reasonable description of proton
distributions (determined from electron scattering experiments)
makes the Gambhir method attractive for predictions
of peripheral neutron distributions for nuclei not yet measured.

\begin{table}[h!]
\caption{Strong interaction level widths and shifts determined by PS209
experiment.}
\resizebox{13.5cm}{!}{
\begin{tabular}{ r l c | l r r | l r r } \hline
Z & A & lower level & 
$\Gamma_{low}$      &$\epsilon_{low}$    &  &
$\Gamma_{up}$       &$\epsilon_{up}$     &   \\
  &   & $n \, , l$  &
 (eV)               &    (eV)            &  &
 (eV)               &    (eV)            &   \\
  &      &          &
                    &      A             & B &
                    &      A             & B  \\ 
\hline
8 & 16O & 3,2 &  484(25) & 103(10) &   &           &  & \\ 
  &     &     &          &         &   &           &  & \\
20& 40Ca& 5,4 &          & 5(12)   &   & 0.059(18) &  & \\ 
  & 42Ca&     &          & 17(14)  &   & 0.080(28) &  & \\
  & 43Ca&     &          & 62(30)  &   & 0.073(42) &  & \\
  & 44Ca&     &          & 31(10)  &   & 0.077(23) &  & \\
  & 48Ca&     &          & 33(12)  &   & 0.116(17) &  & \\
  &     &     &          &         &   &           &  & \\
26& 54Fe& 5,4 & 545(45)  & 155(60) &   & 2.9(6)    &  & \\
  & 56Fe&     & 545(54)  & 167(22) &   & 3.3(5)    &  & \\
  & 57Fe&     & 638(35)  & 164(25) &   & 3.7(4)    &  & \\
  & 58Fe&     & 2017(203)& -115(115) &   & 4.1(10)   &  & \\
  &     &     &          &         &   &           &  & \\
27& 59Co&     & 1370(370)& 254(111)&   & 5.2(7)    &  & \\
  &     &     &          &         &   &           &  & \\
28& 58Ni& 5,4 & 910(150) & 150(40) &   & 4.6(10)   &  & \\
  & 60Ni&     & 1132(173) & 218(62) &  & 5.3(21)   &  & \\  
  & 62Ni&     & 1210(270) & 227(85) &  & 5.6(18)   &  & \\  
  & 64Ni&     & 1570(220) & 349(79) &  & 9.3(18)   &  & \\  
  &     &     &           &          &  &           &  & \\
40& 90Zr& 6,5 & 1040(40)  & 28(29)  &  & 11.5(5)   &  & \\
  & 96Zr&     & 1260(60)  & 169(49) &  & 15.0(17)  &  & \\
  &     &     &           &          &  &           &  & \\
48&106Cd& 7,6 & 199(60)   & -26(20)   &  & 3.8(5)    &  & \\
  &116Cd&     & 251(47)   & -19(16)   &  & 3.0(5)    &  & \\
  &     &     &           &          &  &           &  & \\
50&112Sn& 7,6 & 387(17)   &  -5(11)   &  & 4.2(6)    &  & \\
  &116Sn&     & 382(20)   & 23(13)  &  & 4.9(9)    &  & \\
  &120Sn&     & 474(21)   & 31(13)  &  & 5.6(6)    &  & \\
  &124Sn&     & 512(19)   & 43(11)  &  & 6.1(7)    &  & \\
  &     &     &           &          &  &           &  & \\
52&122Te& 7,6 & 619(59)   & 51(28)  &  & 7.5(10)$^{**}$& -2(8) & -13(8)\\
  &124Te&     & 548(51)   & 51(17)  &  & 7.5(12)$^{**}$& -4(5) & -16(5)\\
  &126Te&     & 651(52)   & 45(28)  &  & 8.4(9)$^{**}$ &5(7) & -9(7)\\
  &128Te&     & 627(60)   & 64(18)  &  & 11.1(17)$^{**}$&15(4)& -8(4)\\
  &130Te&     & 651(119)  & 66(40)  &  & 43.2(86)$^{**}$&-67(4) & -11(4)\\
  &     &     &           &          &  &           &  & \\
70&172Yb& 8,7 & 997(63)   & -422(60)  &202(60)& 31(2)& -9(42)& 134(42) \\
  &176Yb&     & 1121(60)  & -342(36)  &197(36)& 37(2)&10(31)& 133(31) \\
  &     &     &           &          &   &    &      & \\
82&208Pb& 9,8 & 312(26)   & 95(35)  &   & 5.9(8)    &38(18) & \\
83&209Bi& 9,8 & 506(50)   & -8(53)    &   & 6.9(13)  & -14(20) & \\
90&232Th& 9,8 &1534(206)$^*$&-1972(71)$^*$& 42(72)&50.8(72)&-507(30)&74(30)\\
92&238U & 9,8 &2362(221)$^*$&-2832(69)$^*$&335(79)&68.1(61)&-872(16)&-56(20)\\ 
\hline
\multicolumn{8}{l}
{$\epsilon$ -- level shift (positive value means repulsive level shift)} \\
\multicolumn{8}{l}
{A -- measured, B -- corrected for E2 resonance} \\
\multicolumn{8}{l}
{$^*$ -- only lower fine structure level was measured in these cases} \\
\multicolumn{8}{l}
{$^{**}$ -- not corrected for E2 resonance (see Ref.~\cite{klo00})}
\end{tabular}}
\label{all}
\end{table}
Besides the radiochemical method antiprotons offer 
another, more classic way for investigating the nuclear periphery.
The strong interaction level widths and shifts 
in the antiprotonic atoms depend on the antiproton-nucleus interaction 
potential which, in turn, exhibits a strong nuclear density dependence.
During about
750~h of beam time 8.7~$\cdot 10^{10}$ antiprotons were stopped in
55 natural or isotopically separated targets 
from $ ^{16}\mbox{O}$ to $ ^{238}\mbox{U}$. 
At present the
evaluated data consist of 45 level shifts, 29 "lower" level
widths and 33 "upper" level widths, the latter obtained from the
relative yields of X-rays. In addition, for targets with $Z \geq
48$ the fine structure components could in most cases be resolved.
Table~1 gives the summary of the measured level widths and shifts
(for the resolved levels average values are given).

\begin{figure}[h]
\begin{center}
\includegraphics[width=0.6\textwidth]{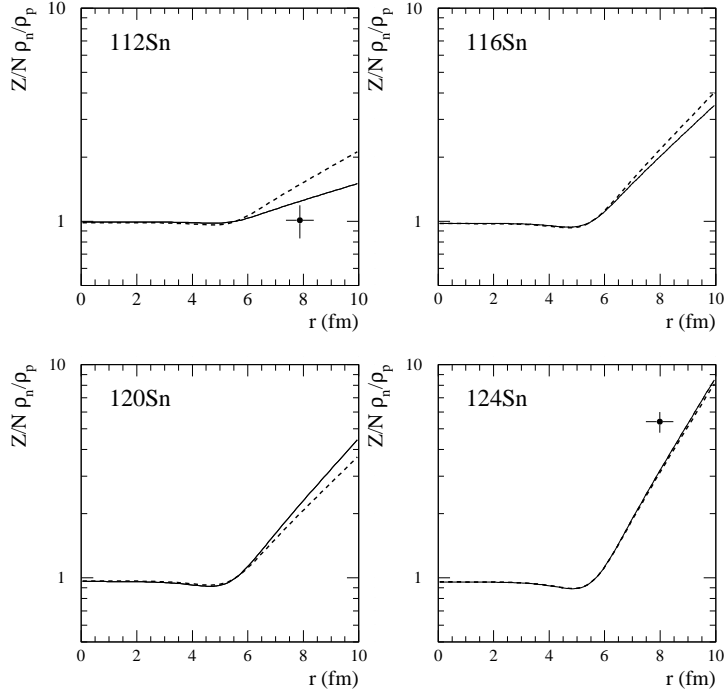}
\end{center}
\caption{Neutron to proton density ratio deduced from strong interaction 
level widths and shifts (dashed lines) for 
$^{112,116,120,124}\mbox{Sn}$~\cite{bat89,kra94,kra99}.
It is comapared with $f_{halo}$ deduced for $^{112}\mbox{Sn}$ 
and $^{124}\mbox{Sn}$ 
(marked with crosses at a radial distance $R_{1/2} + 2.5$~fm).
Modified Batty potential~\cite{bat97} 
with $\frac{Im \, a(\overline{{\rm p}}{\rm n})}
{Im \, a(\overline{{\rm p}}{\rm p})}=0.63$ was used.
The density ratios deduced from $\Delta r_{np}$ 
under the assumption $c_n = c_p$ (see Fig.~\ref{1})
are also shown (solid lines).}
\label{3}
\end{figure}

As demonstrated by the calculation~\cite{wyc96} the antiprotonic X-rays probe
the nuclear periphery at distances about 1 fm closer to the nuclear centre
than the radiochemical method described above. At present  the consistency
of both methods employed is intensively investigated.
Figure~\ref{3} presents neutron to proton density ratio 
for $^{112,116,120,124}\mbox{Sn}$ deduced from the measured level widths and 
shifts. For $^{112}\mbox{Sn}$ and $^{124}\mbox{Sn}$ 
it is compared with the same quantity ($f_{halo}$) 
deduced with the radiochemical method.
In the former method  it was again assumed that 
the proton~\cite{vri87} and neutron densities are given by 2pF distributions.
The agreement between results of these two methods is not good.
A similar discrepancy between the neutron to proton density ratio
deduced from the radiochemical and X-ray experiment were observed
in $^{176}\mbox{Yb}$~\cite{sch98} 
and in $^{128,130}\mbox{Te}$ nuclei~\cite{klo00}.
This maybe caused
by the 2pF distribution not being valid at large radii or  
the $\overline{\mbox{p}}$--nucleus potential used not being adapted
for heavy, neutron rich nuclei.
More elaborate considerations of the antiproton--nucleus interaction
potential are presented during this conference~\cite{wyc00}.

On the other hand, the parameters of the nuclear matter distribution for 
$^{112,116,120,124}\mbox{Sn}$ obtained from the antiprotonic
X-rays measurements give differences $\Delta r_{np}$ between 
neutron and proton root-mean-square radii
which are in good agreement with values determined with other 
methods~\cite{kra94,kra99}. This is seen
in Fig.~\ref{3} where the neutron to proton density ratio 
deduced from $\Delta r_{np}$ measurements and determined from 
our antiprotonic X-ray data for four tin isotopes are shown.

This work was supported by KBN grants 2~P03B~048~15 and~2~P03B~119~16 
and by the Deutsche Forschungsgemeinschaft, Bonn.

\end{document}